\documentclass{article}

\begin{document}

\title{The Modular and Renormalisation Groups in the Quantum Hall Effect
\footnote{Talk presented at the 
{\it Workshop on the Exact Renormalisation Group},
Faro, Portugal, 10-12th September, 1998}}

\author{Brian P. Dolan\\
{\sl Department of Mathematical Physics,}\\
{\sl  National University of Ireland,  Maynooth, Republic of Ireland,}\\
\\ {\sl e-mail: bdolan@thphys.may.ie}}

\maketitle

\abstract{An analytic form for the crossover of the conductivity
tensor between two Hall plateaux, as a function of the external
magnetic field, is proposed. The form of the crossover is obtained
from the action of a symmetry group, a particular subgroup
of the modular group, on the
upper-half complex conductivity plane, by assuming that the
$\beta$-function describing the crossover is a holomorphic
function of the conductivity. The group action also leads to
a selection rule, $\vert p_1q_2-p_2q_1\vert=1$, 
for allowed transitions between Hall plateaux
with filling factors $\nu_1=p_1/q_1$ and  $\nu_2=p_2/q_2$,
where $q_1$ and $q_2$ are odd.
}

\def\autoeql{ {\global\advance\count90 by1} & (\the\count90)}
\def\autoeq{ {\global\advance\count90 by1}\eqno(\the\count90) }
\def\circum#1{{ \kern -3.5pt $\hat{\hbox{#1}}$ \kern -3.5pt}}

\section{The Quantum Hall Effect}

The classical Hall effect occurs when an electric current is
passed through a two dimensional slab of conducting or
semi-conducting material in a perpendicular magnetic field 
${\underline {B}}$.  If the slab lies in the $x-y$ plane, with
its longer axis aligned with the $x$-direction and the
magnetic field in the $z$-direction, then a transverse
voltage, the Hall voltage $V_{xy}$, is generated across the
short axis of the sample, when a current $I$ is passed through
the sample by applying a longitudinal voltage, $V_{xx}$, across
the long axis.  Thus we can define two resistances, a
longitudinal resistance $R_{xx}$ and a transverse resistance
$R_{xy}$ using Ohm's law,
$$
V_{xx} = I \,\,R_{xx}, \qquad V_{xy} = I \,\, R_{xy}.
\autoeq 
$$
In two dimensions resistivity and resistance have the same
dimensions and the classical analysis of the Hall effect
gives the transverse resistivity $\rho_{xy}$ as proportional to
the external field $R_{xy} =  \rho_{xy} = {{B}\over{ne}}$ 
where $n$ is
the density of charge carriers with charge $e$. For homogeneous  conditions the
longitudinal resistivity, $\rho_{xx}$, is related to $R_{xx}$ by a
dimensionless geometrical ratio.  Classically a graph
of $\rho_{xy}$ against $B$ is a straight line whose slope is
inversely proportional to $n$ - giving a useful technique
for measuring $n$ and the sign of $e$ in a semi-conductor.

Quantum mechanically it was discovered by von Klitzing et
al \cite{vK} that, for low enough temperatures and high enough magnetic
fields, $\rho_{xy}(B)$ increases in a series of steps with very
stable plateaux at multiples $1\over \nu$ of a fundamental unit
of resistance $R_H=h/e^2=25.8128k\Omega$, where $\nu$ is an
integer, $h$ is Planck's constant and $e$ the electric
charge.  At the same time, $\rho_{xx}=0$ for values of $B$ 
where $\rho_{xy}={1\over \nu}R_H$.  This is the integer Quantum Hall Effect
(QHE). To achieve some insight into the physics of the integer QHE
consider a 
quantum mechanical two dimensional gas of free
electrons. In an external magnetic field this is equivalent to a
harmonic oscillator with energy levels Landau levels 
$$E_N=\hbar w_c\bigl(N+{1\over 2}\bigr)$$
where $w_c={{eB}\over{m}}$ is the cyclotron frequency and $m$ is
the electron mass --- see for example Landau and
Lifschitz, \cite{LL}.  Each level has a degeneracy, $g$,
proportional to the external magnetic field and the area of
the sample, $A$,
$$g={{eBA}\over{h}}.$$
Defining $n_B={{eB}\over{h}}$ as the degeneracy/unit area we
see that
$$
n_B={{eB}\over{h}}={{ne^2}\over{h}}\rho_{xy},
\autoeq 
$$
and the filling factor, $\nu$, for the Landau levels is
$$
\nu=n/n_B={{h}\over{e^2}} {{1}\over{\rho_{xy}}}
\autoeq
$$
where we had used the \lq classical' expression $\rho_{xy}=B/ne$
to eliminate $B$.  Thus $\rho_{xy} = {1\over \nu}{{h}\over{e^2}}={1\over
\nu}R_H$ and the Hall plateau observed by von Klitzing
correspond to stable states in which $\nu$ Landau levels are
exactly filled.  The stability of these states is analogous
to the chemical stability of the noble gases, when electron
shells are exactly filled.  In order to observe the effect,
one must work at temperatures low enough that
$$
kT \ll \hbar w_c.
$$
For $B\sim 1{Tesla}$, this requires $T\ll 1{K}$.
This simple analysis omits many ingredients, such
as the crucial r\circum{o}le played by impurities in the QHE,
but serves to give some insight into the phenomenon.

Later, in 1982, Tsui et al \cite{Tsui}, observed Hall plateaux at
fractional filling factors $\nu=p/q$ with $p$ and $q$ small
integers, but only with odd $q$.  Again $\rho_{xx}=0$ at the
plateaux.  This is the fractional QHE.  A good review is
Prange and Girvin's book \cite{PrangeGirvin}. Note that the QHE can be
described equally well in terms of conductivities, rather
than resistivities where 
$\sigma = \rho^{-1}, \,\, \sigma = 
\pmatrix{\hfill\sigma_{xx}
&\sigma_{xy}\cr -\sigma_{xy}&\sigma_{xx}}$ and
$\rho = \pmatrix{\hfill\rho_{xx}
&\rho_{xy}\cr -\rho_{xy}&\rho_{xx}}$. 
\section{Scaling and Crossover in the Quantum Hall Effect}

In order to understand how the resistivity changes as $B$ is
varied, one can define $\beta$-functions for $\rho_{xx}$ and
$\rho_{xy}$.  Khmel'nitskii \cite{Khmelnitskii} defined these as
$$\beta_{xx} = L {{d\rho_{xx}}\over{dL}} \quad\quad \beta_{xy} =
L {{d\rho_{xy}}\over{dL}}
\autoeq
$$
where $L$ is a characteristic length for the electron
dynamics --- it would be the mean free path in the absence of
an external magnetic field but in general will depend on the
field. The idea is
that different Hall plateaux can be interpreted as different
phases of the 2-D electron gas and the plateaux should be
fixed points at which the $\beta$-functions vanish. 
Khmel'nitskii suggested a phase diagram for the conductivities,
based on topological considerations for the flow, similar to
the figure below. There is a further fixed point at a
critical field, $B_c$, between any two Hall plateaux ---
where the dotted lines cross the solid lines in the figure.\hfill\break
\bigskip
\hfill\break
\includegraphics{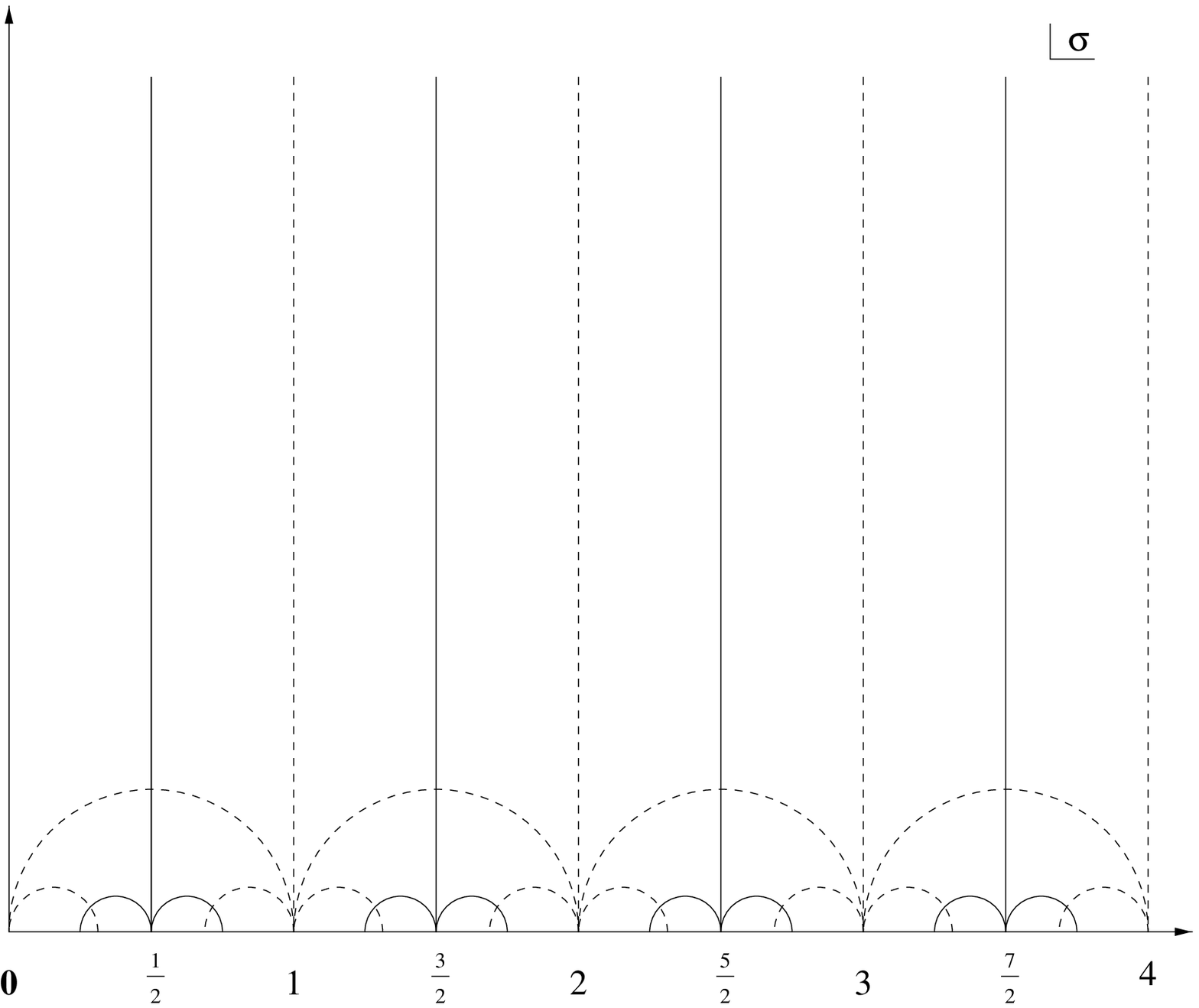}
\vskip 5.5cm
\centerline{\sl Figure 1. The phase diagram for quantum Hall transitions.}
\centerline{\sl Solid lines represent phase boundaries.}
\bigskip
In fact the
transition between two hall plateaux $\nu:p_1/q_1 \rightarrow
p_2/q_2$ is a quantum phase transition. (Fisher \cite{Fisher}). This
differs from the usual phase transitions of statistical
mechanics in that the long range order of the ordered phase is
destroyed, not by thermal fluctuations as the temperature is
varied, but by quantum fluctuations as the external magnetic
field is varied.  Thus there is a correlation length, $\xi$,
which diverges (in an infinite sample) as the external field
approaches a critical value $B_c$ with a critical exponent
$\nu_\xi > 0$, 
$$
\xi \sim\vert B-B_c\vert^{-\nu_{\xi}}.
$$
Pruisken \cite{Pruisken} has argued that only the field $B$ and the
temperature $T$ are relevant near $B_c$, so $L$ should be
a function of $B$ and $T$ only, $L(B,T)$, and similarly for $\rho$. Since,
in units of $h/e^2, \rho_{xy}(B,T)$ is dimensionless it should
only depend on a dimensionless ratio $\tilde b= \Delta B/T^u$
where $\Delta B=B-B_c$ and $\mu$ is an anomalous dimension. At
various temperatures $T$ we can plot $\rho_{xy}\bigl({{\Delta
B}\over{T^\mu}}\bigr)$ as a function of $\Delta B$ and it should
have the same value $\rho_{xy}^{c} = \rho_{xy}(0)$ at
$\Delta B=0$ for all temperatures, though its slope at $\Delta
B = 0$ will be different for different temperatures. 
Experimentally this is how the critical field, $B_c$, for the
transition between two plateaux is measured, see Shahar et al.
\cite{Shaharetal}, figures 1 and 2. There is also experimental evidence, that
$\mu$ is the same for every quantum Hall transition, giving rise
to the concept of super-universality.

Note that, since $\nu \propto 1/B$ classically, we could
equally well use the dimensionless parameter,
$v={{ne}\over{T^\mu}} \Delta\nu$ to describe the crossover
between two Hall plateau, instead of $\tilde b$.

\section{The Modular Group and the Quantum Hall Effect.}

The modular group, $\Gamma(1)$ consisting of $2\times 2$
matrices with integer entries and determinant one,
$Sl(2,{\bf Z})$,
has a natural action on the upper-half complex conductivity
plane $\sigma = \sigma_{xy} + i \sigma_{xx} (\sigma_{xx}\geq
0)$, originally noted by L\"utken and Ross \cite{LutkenRossa},
$$
\sigma\rightarrow {{a \sigma + b}\over{c \sigma + d}}\,,
\qquad\qquad ad-bc=1.
\autoeq
$$
L\"utken and Ross suggested that this should be a
symmetry of the partition function. Fixed points of $\Gamma(1)$ 
should then be fixed points of the renormalisation
group flow. In a further development, \cite{LutkenRossb}, L\"utken and Ross
observed that actually $\Gamma(1)$ is too large ---
it can map filling factors $p/q \rightarrow p^\prime/ q^\prime$ 
with $q$ odd and $q^\prime$ even, which is incompatible
with the experimental observation that only odd denominators
are observed in the QHE.  A better group to use is a
sub-group $\Gamma_0(2) \subset \Gamma(1)$, where
$\gamma\in\Gamma_0(2)$ if
$$
\gamma={a\quad b\choose 2c\quad d},\qquad ad-2bd=1.
$$
$\Gamma_0(2)$ preserves the parity of the denominator. 
Using $\Gamma_0(2)$ one can make predictions about the critical
conductivities, since they should correspond to fixed points
of $\Gamma_0(2)$.  Thus for the transition $\nu :1\rightarrow 2$ one expects
$\sigma_c={{3+i}\over{2}}\Rightarrow\rho_c={{3+i}\over{5}}$,
which predicts that $\rho_{xy}^c=3/5, \rho_{xx}^c=1/5$ which
can be compared with the experimental results in \cite{Shaharetal}.

The assumption that $\Gamma_0(2)$ is a symmetry also allows
one to derive a selection rule for allowed quantum Hall
transitions $\nu:p_1/q_1\rightarrow p_2/q_2,$ \cite{Modular}. By
assumption $\nu:p_1/q_1\rightarrow p_2/q_2$ can be obtained
from $\nu: 0\rightarrow 1$ by some $\gamma={a\quad b\choose
2c\quad d}\in \Gamma_0(2)$. Now 
$\gamma(0)={b\over d} ={{p_1}\over{q_1}}$, 
$\gamma(1)={{a+b}\over{2c+d}}=p_2/q_2$, so
$\gamma = {\pmatrix{p_2-p_1&p_1\cr q_2-q_1&q_1\cr}}$. The
condition $det\gamma =1$ now requires $p_2q_1-p_1q_2=1$,
which is a selection rule which is well observed in the
experimental data, see for example \cite{Willetetal}.  
The critical conductivity for the
transition is predicted by $\Gamma_0(2)$ symmetry to be
$$
\sigma_c =
\gamma\biggl({{1+i}\over{2}}\biggr)={{p_2q_2+p_1q_1+i}\over{(q_1^2+q_2^2)}}
\autoeq
$$
in units of $e^2/h$. This action of $\Gamma_0(2)$ is in fact an
extension to the complex conductivity plane of the
\lq\lq law of corresponding states'' of reference \cite{KLZ}.

\section{Crossover in the Quantum Hall Effect}

By making certain assumptions about the analytic form of the
crossover, compatible with $\Gamma_0(2)$ symmetry, one can
make very strong predictions which can then be used to test
the assumptions.  We shall define $\beta$-functions for
$\sigma$ using the dimensionless variable $v$ defined
previously, but in order to be as general as possible we
shall define
$$
\beta={{d\sigma}\over{ds}} \autoeq
$$
where $s$ is some real analytic monotonic function of $v$. In
general, $\beta$ might depend on both $\sigma$ and its
complex conjugate, $\bar \sigma$, but if we make the
assumption that $\beta$ is analytic (holomorphic) 
and depends only on
$\sigma$ we get some very tight restrictions on the form of
$\beta$ (a possible form of the beta-function with is not
holomorphic was proposed in \cite{BurgessLutken}).\footnote{After this
talk was presented at Faro another
non-analytic $\beta$-function was proposed in \cite{Tanaguchi}.}
Since
$$
\beta(\gamma(\sigma)) =
{{d\gamma(\sigma)}\over{dS}}={{1}\over{(2c\sigma+d)^2}}\beta(\sigma)\autoeq$$
$\beta$ is what is known in the mathematical literature as a
modular form of weight $-2$ for the group
$\Gamma_0(2)$ (See e.g. Ran-kin \cite{Rankin}).  We shall now follow
an analysis similar to that of Ritz \cite{Ritz}.  Such modular
forms can be written as a ratio of two polynomials
$$
\beta(\sigma)={{c}\over{f^\prime}}
\mathop{\Pi}\limits_{i=1}^{N}(f-\alpha_i)^{m_i}\autoeq
$$
where $m_i\in{\bf Z}$ and $c$ and $\alpha_i$ are constants.  The function
$f(\sigma)$ is defined to be 
$f(\sigma)= -{{\theta_3^4\theta_4^4}\over{\theta_2^8}}$ where
$\theta_2=\sum\limits_{n}^{} e^{\pi i(n+1/2)^2\sigma},
\theta_3=\sum\limits_{n}^{}e^{\pi in^2\sigma},
\theta_4=\sum\limits_{n}^{}(-1)^n e^{\pi in^2\sigma}$ are
Jacobi $\theta$-functions, and $f^\prime =
{{df}\over{d\sigma}}$.  $f$ has the property of being
invariant under $\Gamma_0(2)$, $f(\gamma(\sigma)) =
f(\sigma)$, as shown in \cite{Rankin}. 

We can obtain information about the possible values of $\alpha_i$
by appealing to experiment. By assumption 
the $\beta$-function should
vanish at $\sigma=1$ and $\sigma = 2$, since these are are
Hall plateaux.  Equivalently, since 
$\sigma\rightarrow \sigma-1$ is a  $\Gamma_0(2)$ transformation, the
$\beta$-function should vanish at $\sigma = 0$ and $\sigma =1$.
Experimentally \cite{Shaharetal} there is also a critical  point at 
$\sigma_c={{1+i}\over{2}}$ (see
figure 2).  Now analytically $f(0)=f(1)=0$ and
$f({{1+i}\over{2}})=1/4$, therefore $\alpha_i=0$ or 1/4.  We shall
make the minimal assumption that there are no other critical
points.  Thus
$$
\beta(\sigma) = {{c}\over{f^\prime}} f^n(f-1/4)^m
\autoeq $$
for some integers $n$ and $m$ and a constant, $c$.  Further
constraints can be placed on $\beta$ by making some further,
rather reasonable, assumptions.
\begin{itemize}
\item{$\beta$ is finite as $\sigma\rightarrow i \infty
\Rightarrow n+m \leq 1$ (since $f\rightarrow -\infty$ 
and $f^\prime \rightarrow -2\pi i f$ as $\sigma \rightarrow i\infty$).}
\item{$\beta\rightarrow 0$ as $\sigma \rightarrow 0$
$\Rightarrow n \geq 1$, and so $m\leq 0$ (since 
$f\rightarrow -16\hbox{e}^{-{i\pi\over\sigma}}$ and $f^\prime\rightarrow i\pi {f\over\sigma^2}$ as $\sigma\rightarrow 0$, with positive imaginary part).}
\item{${{d\sigma_{xx}}\over{ds}}\leq 0$ and
${{d\sigma_{xy}}\over{ds}}=0$ as $\sigma\rightarrow i
\infty\Rightarrow c$ is real.}
\end{itemize}
Without loss of generality,
we can choose $c=1$. This leads to the form
$$
\beta(\sigma)={{{(-1)}^{\tilde m+n}}\over{f^\prime}}{{f^n}\over{(f-1/4)^{\tilde m}}}
\autoeq$$
where $\tilde m=-m \geq 0$ and $1\leq n \leq 1 +
\tilde m$ (further details are given in \cite{Crossover}).  The simplest
case is $\tilde m=0, n=1$ and there are arguments \cite{Crossover} that
this is the best choice compatible with experiment, because
it gives the fastest approach to Hall plateaux.  One then has
$$
{{d\sigma}\over{ds}} = - f/f^\prime
\autoeq$$
which was considered in a completely different context, that
of QCD, by Latorre and L\"utken in \cite{LatorreLutken}.  The resulting
renormalisation group flow can be determined by integrating
$$
{{ds}\over{d\sigma}}= - f^\prime /f \Rightarrow s(\sigma) =
- \ln f/f_0 +s_0
\autoeq$$
where $s_0$ is a constant and $f_0=f(\sigma(s_0))$. Since
$s$ is real, $f/f_0$ must be real and thus $arg(f)$
is constant along the trajectories.  The flow is shown in
the figure below, which is simply a contour plot of $arg(f)$, and
reproduces the topology of Khmel'nitskii's flow diagram, but with the vertical axis normalised so that there are unstable fixed points at 
$\sigma_c={1+i\over 2}$ and its images under $\Gamma_0(2)$.\hfill\break
\smallskip\hfill\break
\includegraphics{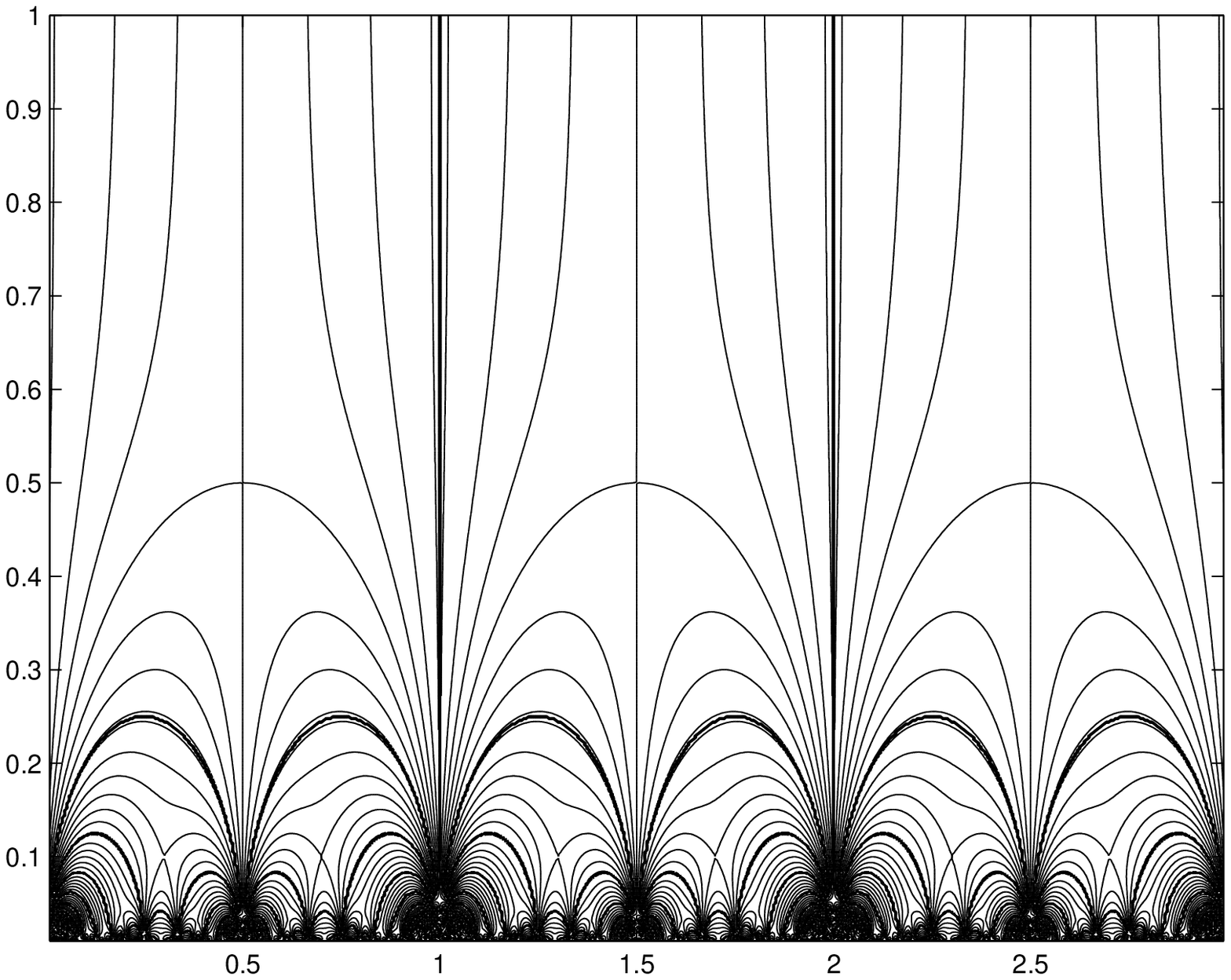}
\vskip 4.8cm
\centerline{\sl Figure 2. The flow diagram for the complex conductivity}
\bigskip
We can obtain an
explicit form for $\sigma(s)$ using well known relations
between Jacobi $\theta$-functions and complete elliptic
integrals of the second kind
$$
K(k) = \int_0^{\pi_{12}}
{{d\phi}\over\sqrt{1-k^2\sin^2\phi}},
\autoeq$$
see for example Whittaker and Watson \cite{WW},
$$
\theta_2^2 = {{2kK(k)}\over{\pi}} \quad 
\theta_3^2 = {{2K(k)}\over{\pi}} \quad 
\theta_4^2 = {{2k^\prime K(k)}\over{\pi}}$$
where $e^{i\pi\sigma}=e^{-\pi K^\prime(k)/K(k)}$ and
$K^\prime(k) = K(k^\prime)$ with $(k^\prime)^2 = 1-k^2$ the
complimentary modulus.  Using these relations, and equation 
(13), one can show that, for the transition 
$\nu :0\rightarrow 1$,
$$
\sigma(s)= {{K^\prime(w)\{K^\prime(w)+iK(w)\}}
\over{(K(w))^2+(K^{\prime}(w))^2}}
\autoeq$$
where $w^2 = {{1\pm\sqrt{1-e^{-s}}}\over{2}}$,
with $s>0$ ($s<0$ corresponds to the vertical lines through
$\sigma_c={n+i\over 2}$ in figure 2, which
are irrelevant directions ---details
are given in \cite{Crossover}).  

To make contact with experiment, we must say something about
how $s$ depends on $\Delta \nu$ or, equivalently, $\Delta
B$. Experimentally the slope of $\sigma_{xy}$ as a function
of $\Delta\nu$ is finite and non-zero at $\Delta\nu=0$,
(i.e. at $\sigma_c={{1+i}\over{2}}$).  Analytically one
finds $\beta(\sigma)\propto{{1}\over\sqrt{s}}$ near $s=0$.
If one takes $s\propto\Delta\nu^2$, 
$s=({{A\Delta\nu}\over{T^\mu}})^2$ 
near $\Delta\nu=0$ where $A$ is a constant,
then ${{d\sigma}\over{d(\Delta\nu)}}$ is perfectly finite at
$\Delta\nu=0$.  Equation (15) can now be used to determine
$\sigma$ for the crossover between any Quantum Hall plateau
$\nu : p_1/q \rightarrow p_2/q_2$, by acting on it with an
element, $\gamma$, of $\Gamma_0(2)$.  A subtlety is that one
should also transform $\nu$ by the {\it inverse} $\gamma^{-1}$.
The final result is
$$
{\sigma ={{p_1q_1(K(w))^2+p_2q_2(K^\prime(w))^2+iK(w)K^\prime(w)}
\over{q_1^2(K(w))^2+q_2^2(K^\prime(w))^2}}}$$
$$w^2 ={{1-\hbox{sign}\,\,(\Delta\nu)
\sqrt{1-e^{{-({A\Delta\nu\over\zeta T^\mu})^2}}\over{2}}}}
\autoeq$$
where $\zeta(\Delta\nu)=\alpha\{(q_1-q_2)\Delta\nu +\alpha\}$ with
$\alpha = (p_2-p_1)-(q_2-q_1)\nu_c$. A plot of
$\sigma(\Delta\nu)$ for the transition $\Delta\nu :1\rightarrow 2$ 
is given in figure 3, which shows
remarkable argument with the experimental data in figure 1
of reference \cite{Shaharetal}. The plot was obtained by using the experimental value
$\mu=0.45$ and the choice $A=55$ which appears to give a good fit to the data in
\cite{Shaharetal}.
\vfill\eject
\includegraphics{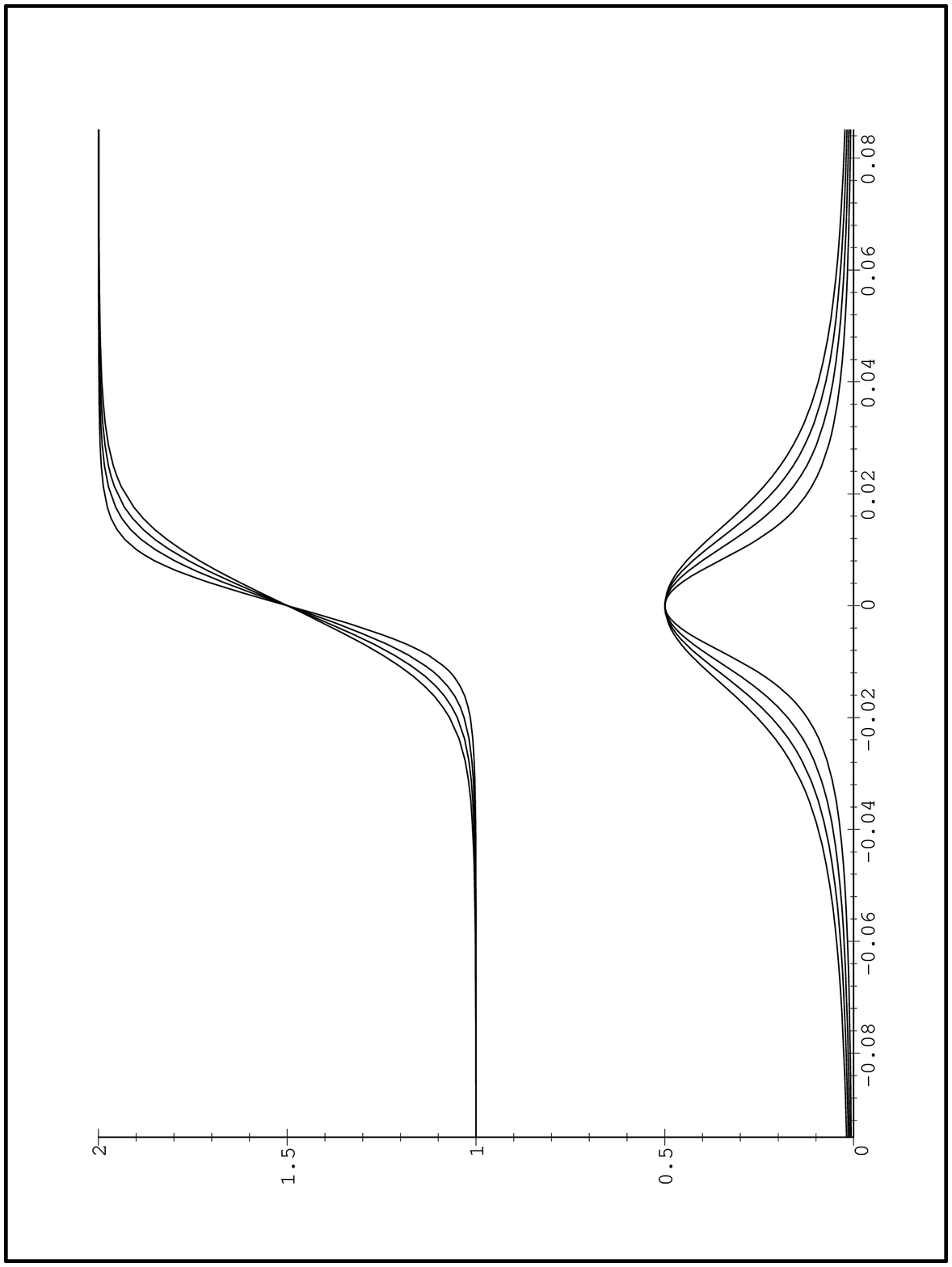}
\centerline{}
\vskip 3cm
$\sigma$
\vskip 3cm
\hskip 4.3cm $\Delta \nu$
\vskip 0.6cm
{\sl \centerline{Figure 3: Crossover of conductivity for $\nu:1\rightarrow 2$
at the four temperatures}
\centerline{$T=42$, $70$, $101$ and $137mK$ with $A=55$ and $\mu=0.45$.
To be compared }
\centerline{with the experimental data in figure 1 of {\cite{Shaharetal}}.} 
\vfill\eject

\end{document}